\documentclass[pra,showpacs,floatfix,preprint,amsmath,amssymb]{revtex4}

\usepackage{epsfig}
\usepackage{amsmath}
\usepackage{supertabular}
\usepackage{graphicx}
\usepackage{dcolumn}
\usepackage{bm}
\usepackage{float}
\usepackage{subfigure}
\usepackage{color}
\begin{document}


\title{Distribution of quantum coherence and quantum phase transition in the Ising system}

\author{Meng Qin$\footnote{E-mail
address: qrainm@gmail.com}$}
\affiliation{
Department of General Education, Army Engineering University of PLA, Nanjing 211101, China}

\noindent

\begin{abstract}
Quantifying of quantum coherence of a given system not only plays an important role in quantum information science but also promote our understanding on some basic problems, such as quantum phase transition. Conventional quantum coherence measurements, such as $l_{1}$-norm of coherence and relative entropy of coherence, has been widely used to study quantum phase transition, which usually  are basis-dependent. The recent quantum version of the Jensen¨CShannon divergence meet all the requirements of a good coherence measure. It is not only a metric but also can be basis-independent. Here, based on the quantum renormalization group method we propose an analysis on the critical behavior of two types Ising systems when distribution of quantum coherence. We directly obtain the trade-off relation, critical phenomena, singular behavior, and scaling behavior for both quantum block spin system. Furthermore, the monogamy relation in the multipartite system is also studied in detail. These new result expand the result that quantum coherence can decompose into various contributions as well as enlarge the applications in using basis-independent quantum coherence to reflect quantum critical phenomena.
\end{abstract}

\maketitle

\section{\label{sec:level1}INTRODUCTION}
 Quantum coherence originating from quantum superposition is the distinguished feature in quantum mechanics. Based on the framework of resource theory, Baumgratz \textit{et al}. \cite{baumgratz} proposed that quantum coherence also can be viewed as one useful quantum resource like entanglement, quantum discord, and so on. It is believed to provide advantages for a variety of informational tasks such as nanoscale thermodynamics \cite{narasimhachar}, quantum metrology \cite{demkowicz}, computational task  \cite{deutsch}, quantum cryptography \cite{gisin}, even in quantum biology \cite{licm,yus,kok,fanzy}. Quantum coherence has attracted a great deal of attention as it plays an essential role in quantum information processing tasks. Many remarkable efforts have been dedicated to study the quantification of quantum coherence.  For instance, the relative entropy of coherence \cite{baumgratz}, the $l_{1}$ norm of coherence \cite{baumgratz}, the Tsallis relative $\alpha$ entropies \cite{rastegin}, the entanglement-based measure of coherence \cite{streltsov}, trace-norm distance \cite{shaolh}, skew information \cite{girolami}, the convex roof measure of coherence \cite{yuanx}, and the robustness of coherence are fully investigated \cite{huml}.  Among them, one kind of measure named square root of the quantum Jensen-Shannon divergence (QJSD) \cite{radhakrishnan} attracted much attention. It is not only a well-defined measure but, also, it satisfy the triangle inequality and symmetric. Radhakrishnan \textit{et al.} \cite{radhakrishnan} give the definition of such quantum coherence, and then examine the distribution in multipartite systems. They pointed out that multipartite coherence can decompose into local parts and intrinsic parts, and there are trade-off relations between them.

In recent years, many researchers are devoted to investigate quantum phase transition (QPT), i.e., the ground state of a many-body system will have abrupt changes at or near zero temperature when one or more parameters are varied. Contrary to thermal phase transition, QPT is driven purely by quantum fluctuations. The relation between QPT and quantum coherence also has deserved extensive investigations. Karpat, \textit{et al}. found that the single-spin coherence can estimate the critical point in the anisotropic XY chain \cite{karpat2014}. Malvezzi \textit{et al}. showed that single site coherence can successfully detect the Ising-like second order phase transition in spin-1 Heisenberg chain \cite{malvezzi}. Qin \textit{et al.} proven that $l_{1}$ norm coherence can probe the QPT in the XXZ model under dynamics condition \cite{qin}. Apart from the above study, the other related literature can be seen in Ref. \cite{renj, zhanggq, chenjj, chandrashekar2017,liyc2016,mzaouali}. Although some progress have been made, it is, however, not known the role of multipartite coherence in detecting QPT when it decomposed into local and intrinsic parts, nor to guarantee the scaling behavior of local coherence and intrinsic coherence when quantum renormalization group (QRG) are introduced. The intrinsic connections between QPT and quantum coherence may need further discussion under QRG.

In this paper, relating QRG method, we study an interesting question that how to use QJSD and its distribution to reflect quantum critical phenomena in many-body system. Renormalization group are a way to deal with problems involving multiple length scales. It main purpose is to reduce the effective degrees of freedom of the system. Through an iterative procedure a mathematically manageable situation is reached eventually. QRG is divided to two steps. The first step is the  kadanoff's transformation and the second steps rescales the size of the system. The major purpose in this study is to investigate the intrinsic connections between QPT and quantum coherence. This will give two perspectives on (i) how QJSD and its distribution evolves as the size of the system becomes large and (ii) the connects of the nonanalytic behavior of QJSD to the critical phenomenon of the system \cite{kargarian}. Based on the tranverse field Ising model, we present some results for above problems. The non-analytic property and scaling behavior is investigated too. Furthermore, we also discuss the effect of Dzyaloshinskii-Moriya (DM) interaction on the distribution of quantum coherence. The DM interaction arising from the spin-orbit coupling can influence the phase transition and the critical behavior of many models \cite{mafw}.

This paper is organized as follows. In the next section, we will introduce the concepts of total, local and collective quantum coherence. In Sec. III, the  one-dimensional Ising model and the distribution of quantum coherence at critical point are given. In Sec. IV, the distribution of quantum coherence in one-dimensional Ising model with DM interaction near the critical point are presented and the monogamy relation are given. The last section is a summary of our study.
\section{\label{sec:level2}Basis-independent OF TOTAL, LOCAL, AND COLLECTIVE QUANTUM COHERENCE}
In 2014, Baumgratz \textit{et al}. \cite{baumgratz} first proposed two kinds of quantum coherence measures that is $l_{1}$-norm of coherence and relative entropy of coherence. But different with entanglement or quantum discord, both methods are basis-dependent. So there are some confusion on this issue. In 2019, Radhakrishnan \textit{et al}. \cite{radhakrishnan2} introduced a basis-independent measure named QJSD to overcome the difficulties. For two quantum states $\rho$ and $\varrho$ $\in \mathcal{B}(H_{1}^{+})$, the QJSD is defined as
\begin{equation}
\mathcal{T}(\rho,\varrho)=\frac{1}{2}[S(\rho\|\frac{\rho+\varrho}{2})+S(\varrho\|\frac{\rho+\varrho}{2})]
=S(\frac{\rho+\varrho}{2})-\frac{S(\rho)+S(\varrho)}{2}.
\end{equation}
where $S(\theta)=-Tr (\theta \log \theta)$ is the von-Neumann entropy. The QJSD measures the distance of two states, but it does not obey the triangle inequality and therefore is not a metric. However, the square root of QJSD
\begin{equation}
C(\rho,\varrho)=\sqrt{\mathcal{T}(\rho,\varrho)}.
\end{equation}
not only meet all the requirements of a good quantum coherence measure needs but also is a metric. We here use its as the total quantum coherence. In Ref. \cite{radhakrishnan2} the authors attribute the total coherence to two different contributions that is local coherence and collective coherence,
\begin{equation}
C\leq C_{l}+C_{c}.
\end{equation}
The local coherence $C_{l}$ is the coherence contained in each subsystem of a quantum state,
the basis-independent local coherence is defined by
\begin{equation}
C_{l}(\rho)\equiv D(\pi^{min},\rho_{I})=\sqrt{S(\frac{\pi_{\rho}+I/d}{2})-\frac{S(\pi_{\rho})+\log_{2}d}{2}}.
\end{equation}
where $\pi^{min}=\pi_{\rho}=\rho_{1}\bigotimes\rho_{2}\bigotimes\cdots \rho_{N}$,
$\rho_{i}$ is the reduced density matrix for the $i$th subsystem, $I$ is the identity matrix in a $d$ dimensional
Hilbert space.

The collective coherence comes from the collective participation of several subsystems and defined as
\begin{equation}
C_{c}(\rho) =\sqrt{S(\frac{\rho+\pi_{\rho}}{2})-\frac{S(\rho)+S(\pi_{\rho})}{2}}.
\end{equation}
where $\pi_{\rho}$ are identical with eq. (4).

\section{\label{sec:level2}DISTRIBUTION OF QUANTUM COHERENCE IN THE TRANVERSE-FIELD ISING MODEL}

\subsection{Renormalization of tranverse-field Ising model}

The Hamiltonian of a one-dimensional tranverse-field Ising model reads \cite{kargarian,liucc,qin2}
\begin{equation}
H=-J\sum_{i=1}^{N}(\sigma_{i}^{z}\sigma_{i+1}^{z}+g\sigma_{i}^{x}),
\end{equation}
where $J$ is the exchange interaction,  $\sigma^{\tau} (\tau=x,z)$ is standard Pauli operators at site $i$, $g$ represents the transverse field. We select two-site as one block. Such two-site blocks can be viewed as one-site in the renormalized subspace.
\begin{figure}[!h]
\centering
\includegraphics[scale=0.5]{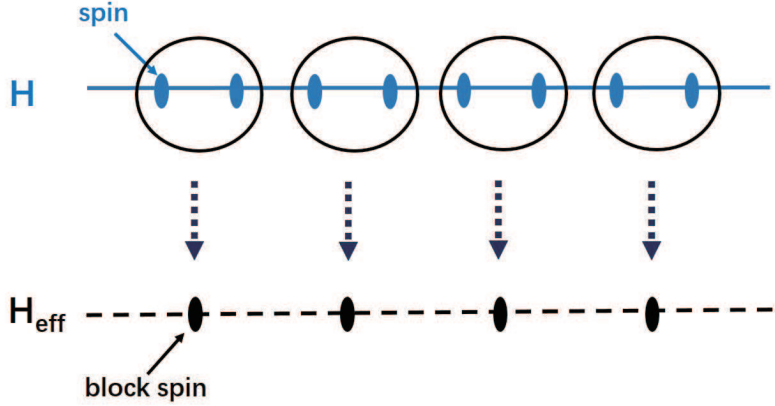}
\caption{(Color online) A schematic description of QRG for two-site as one block.}
\label{fig1}
\end{figure}

After separating the whole systems into two-site blocks, the Hamiltonian can be divided as block Hamiltonian $H^{B}$ and interacting Hamiltonian $H^{BB}$ respectively,
 \begin{equation}
H^{B}=-J\sum_{I=1}^{N/2}(\sigma_{1,I}^{z}\sigma_{2,I}^{z}+g\sigma_{1,I}^{x}),
\end{equation}
\begin{equation}
H^{BB}=-J\sum_{I=1}^{N/2}(\sigma_{2,I}^{z}\sigma_{1,I+1}^{z}+g\sigma_{2,I}^{x}),
\end{equation}

The two degenerate ground state of the corresponding $I$-th block are

 \begin{equation}
\begin{aligned}
|\psi_{0}\rangle=\alpha|00\rangle+\beta|11\rangle,\\
|\psi_{1}\rangle=\alpha|01\rangle+\beta|10\rangle,
\end{aligned}
\end{equation}
where $\alpha=s/\sqrt{s^{2}+1}$, $\beta=1/\sqrt{s^{2}+1}$, and $s=\sqrt{g^{2}+1}+g$.
 In order to get the critical properties of the system at absolute zero,
  we eliminate the excited state by integral and only retain the ground state parts.
 The projection operator are built for this aim. The relations between the
 original Hamiltonian and the effective Hamiltonian is associated by the projection operator, which
 is constructed by two lowest eigenstates
 \begin{equation}
T=\prod_{i=1}^{N/2}T^{L}=\prod_{i=1}^{N/2}(|\Uparrow\rangle _{L}\langle \psi|+|\Downarrow\rangle _{L}\langle \psi|),
\end{equation}
where $\langle\Uparrow|, \langle\Downarrow|$ are renamed states of each block to
represent the effective site degrees of freedom. $|\psi\rangle$ is the
 ground state $|\psi\rangle_{0}$ or $|\psi\rangle_{1}$ .
 The effective Hamiltonian is defined by
\begin{equation}
H_{eff}=T^{\dagger}HT=H^{0}_{eff}+H^{1}_{eff}=T^{\dagger}H^{B}T+T^{\dagger}H^{BB}T.
\end{equation}
The form of effective Hamiltonian are similar to the original Hamiltonian,
\begin{equation}
H_{eff}=-J'\sum_{I=1}^{N/2}(\sigma_{I}^{z}\sigma_{I+1}^{z}+g'\sigma_{I}^{x}),
\end{equation}
where the renormalized couplings are
\begin{equation}\label{eq11}
J'=J\frac{2(\sqrt{g^{2}+1}+g)}{1+(\sqrt{g^{2}+1}+g)^{2}},~~~~~
g'=g^{2}.
\end{equation}

Following above renormalization equation, the critical point $g_{c}$ can be derived by solving $g'=g^{2}$. After some algebra, we can get the nontrivial fixed point $g_{c}=1$. When  $g>1$, the system is in the paramagnetic phase. When $0<g<1$, the system is in the long-ranged ordered Ising phase.

The ground state density matrix is given by
\begin{equation}
\rho=|\psi_{0}\rangle\langle\psi_{0}|=\left(
  \begin{array}{cccc}
    \beta^{2} & 0 & 0 & \alpha\beta \\
    0 & 0 & 0 & 0 \\
    0 & 0 & 0 & 0 \\
    \alpha\beta & 0 & 0 & \alpha^{2} \\
  \end{array}
\right)
\end{equation}
where $\alpha$ and $\beta$ are identical with above results.






\subsection{The critical behavior of quantum coherence}

\begin{figure}[!h]
\centering
\includegraphics[scale=0.5]{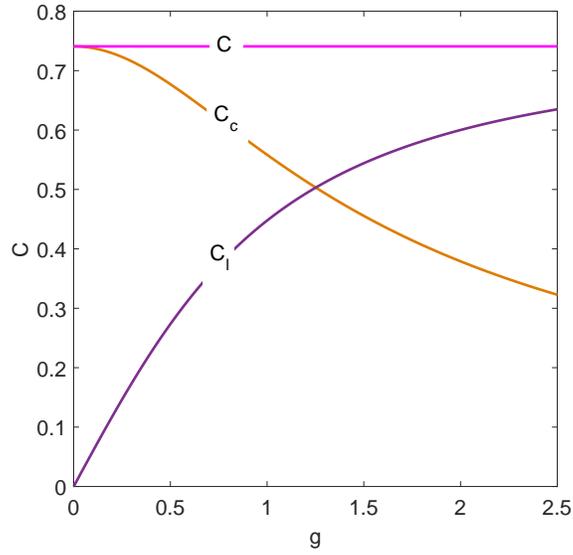}
\caption{Total coherence $C$, local coherence $C_{l}$, and collective coherence $C_{c}$ of tranverse-field Ising systems as the function of $g$.}
\label{fig2}
\end{figure}
Now we investigate the quantum coherence in the two-site tranverse-field Ising systems. We numerically study the dependence of total quantum coherence $C$, collective coherence $C_{c}$ and local coherence $C_{l}$ of the model on the magnetic field $g$ in Fig.~\ref{fig2}. The total quantum coherence can be decomposed of the collective coherence between the two block spin, and the local coherence in each block spin. With the parameters $g$ changes, there are trade-off relations between the two kinds different coherence contributions. This proves that such trade-off relation also can exist in the quantum block state.
\begin{figure}[!h]
\centering
\includegraphics[scale=0.5]{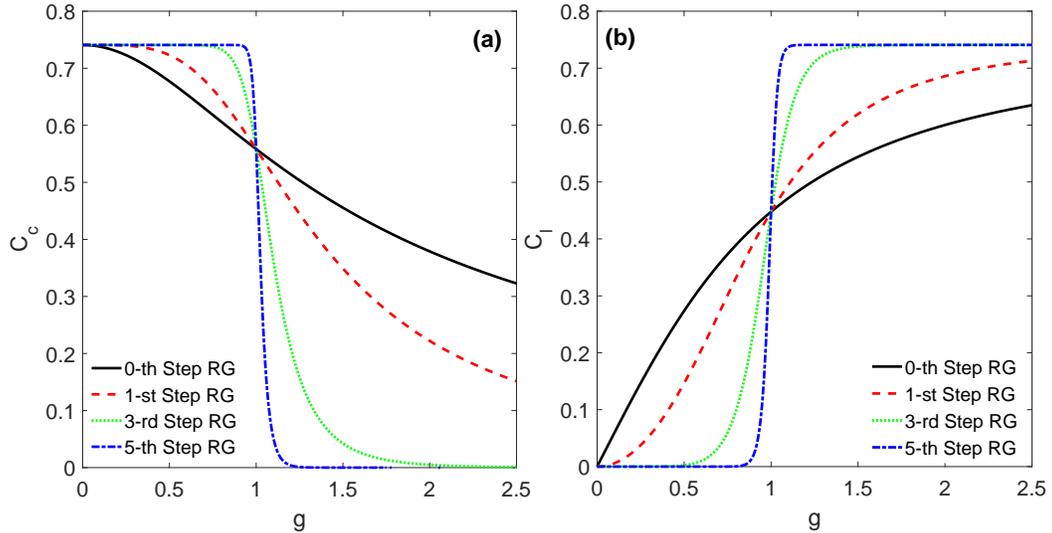}
\caption{Collective coherence $C_{c}$ and local coherence $C_{l}$ of tranverse-field Ising systems as the function of $g$ at different QRG iteration steps.}
\label{fig3}
\end{figure}

The capable of the local and collective quantum coherence in the ITF model in detecting the second-order QPT is displayed in Fig.~\ref{fig3}. The local and collective quantum coherence will develop two different saturated values: $C_{c}=0.7408$ for $0\leq g <1$ and $C_{c}=0$ for  $g>1$, while $C_{l}=0$ for $0\leq g <1$ and $C_{c}=0.7408$ for  $g>1$. This result reconfirmed the trade-off relationship in the block state. In Fig.~\ref{fig3}, the quantum critical point is captured by iterative renormalization steps which manifests that the local and collective quantum coherence can be adopted as an indicator of QPT.

\subsection{Non-analytic behavior and scaling behavior}
\begin{figure}[!h]
\centering
\includegraphics[scale=0.5]{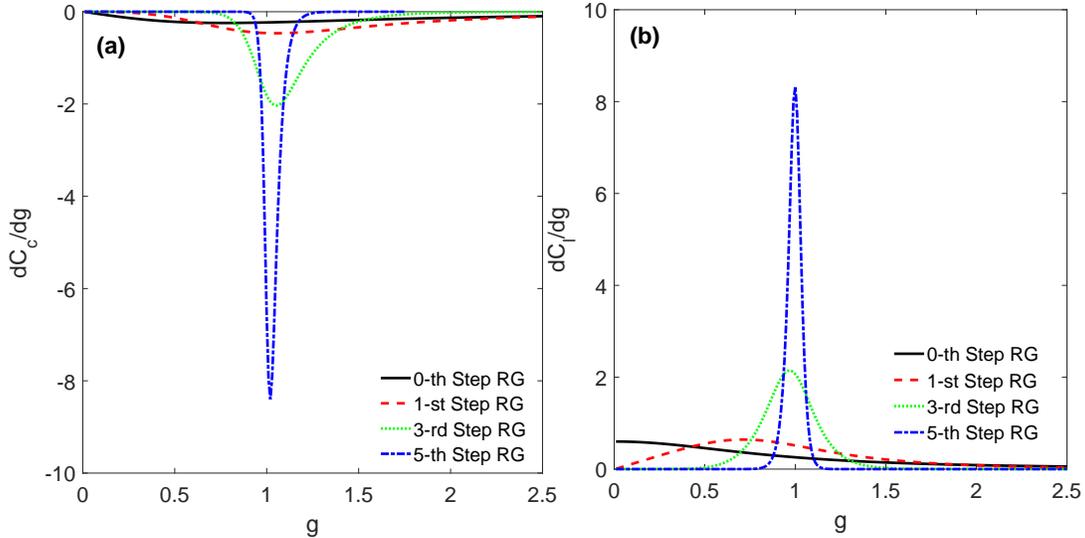}
\caption{The first derivative of collective coherence $C_{c}$ and local coherence $C_{l}$ of tranverse-field Ising systems as the function of $g$ at different QRG iteration steps. }
\label{fig4}
\end{figure}

\begin{figure}[!h]
\centering
\includegraphics[scale=0.5]{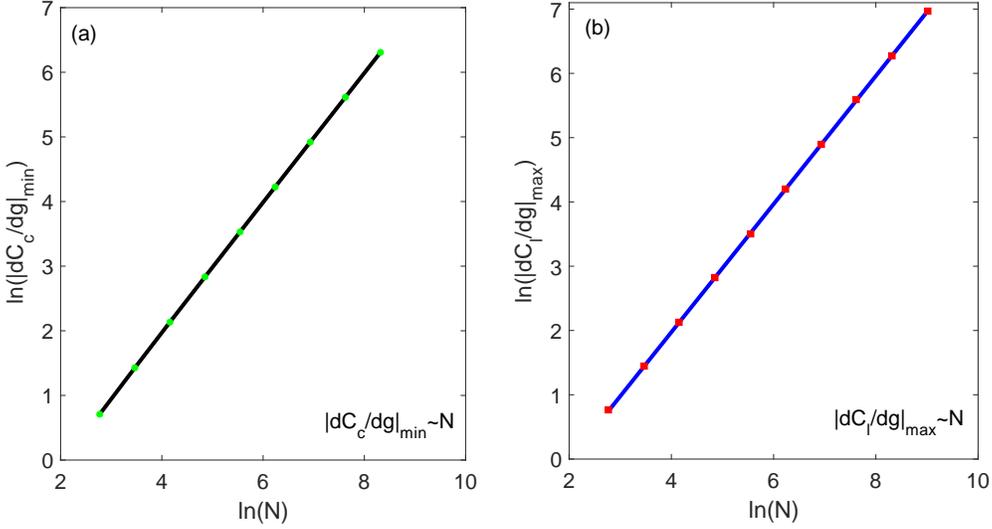}
\caption{The scaling behavior of $\ln |dC_{c}/dg|$ and $\ln |dC_{l}/dg|$ in terms of system size $\ln(N)$.}
\label{fig5}
\end{figure}
In Fig.~\ref{fig4}, we show the first derivative of collective coherence $C_{c}$ and local coherence $C_{l}$ as a function of $g$.
The collective coherence and local coherence divergence as iterative steps increases, which verify that the second-order QPT \cite{leila} happened at the $g_{c}=1$ point. The scaling behavior of the logarithm of collective coherence $\ln |dC_{c}/dg|$ and local coherence $\ln |dC_{l}/dg|$  versus the system size $\ln(N)$ are given in Fig.\ref{fig5}. The collective coherence and local coherence all show scaling behavior when they near the quantum critical point. The scaling law for a finite-size is $|dC_{c}/dg|_{min}\sim N$ and $|dC_{l}/dg|_{max}\sim N$. The quantum critical exponents $\theta$ are correlated with the correlation length exponent $\nu$. The scaling law in Fig. \ref{fig5} shows $\theta=1$, which is equal with the exact result.

\section{\label{sec:level3}DISTRIBUTION OF QUANTUM COHERENCE IN THE ISING MODEL WITH DZYALOSHINSKII-MORIYA INTERACTION}
The one-dimensional Ising model with DM interaction in the $z$ direction is described by the Hamiltonian \cite{jafari2008}
\begin{equation}
H=\frac{J}{4}[\sum_{i=1}^{N}\sigma_{i}^{z}\sigma_{i+1}^{z}+D(\sigma_{i}^{x}\sigma_{i+1}^{y}-\sigma_{i}^{y}\sigma_{i+1}^{x})],
\end{equation}
where $J$ denotes the exchange interaction, $D$ represents the DM interaction, and $\sigma$ is the usual Pauli operators.

\begin{figure}[!h]
\centering
\includegraphics[scale=0.5]{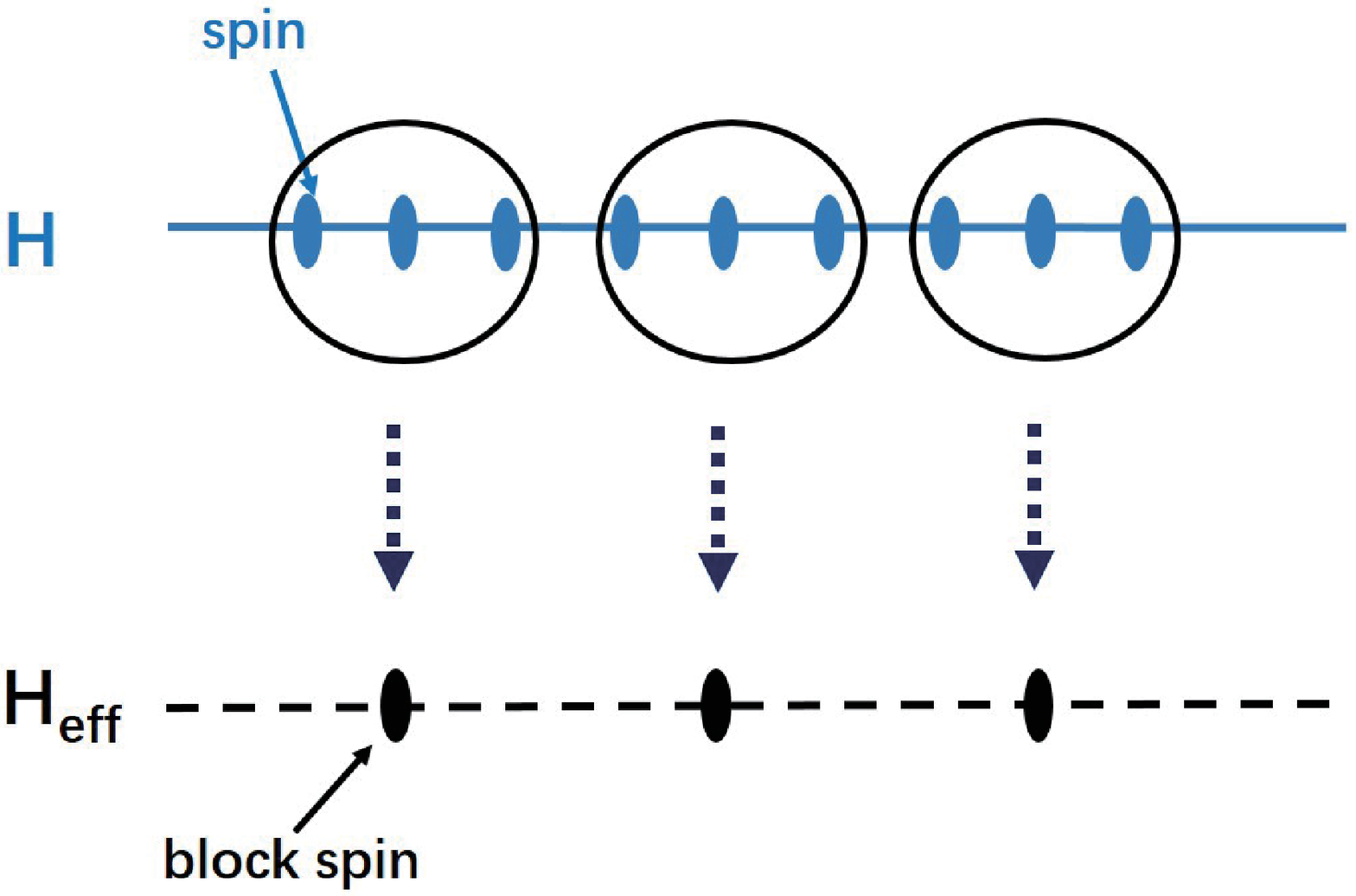}
\caption{(Color online) A schematic description of QRG for three-site as one block.}
\label{fig6}
\end{figure}

A three-site block procedure have considered in this model, defined in Fig. \ref{fig6}. According to the renormalization group method, the Hamiltonian can decompose into interblock $H_{BB}$ Hamiltonian

\begin{equation}
H_{B}=\frac{J}{4}\sum_{I=1}^{N/3}[\sigma_{1,I}^{z}\sigma_{2,I}^{z}+\sigma_{2,I}^{z}\sigma_{3,I}^{z}
+D(\sigma_{1,I}^{x}\sigma_{2,I}^{y}-\sigma_{1,I}^{y}\sigma_{2,I}^{x}+\sigma_{2,I}^{x}\sigma_{3,I}^{y}-\sigma_{2,I}^{y}\sigma_{3,I}^{x})],
\end{equation}
and intrablock $H_{B}$ Hamiltonian

\begin{equation}
H_{BB}=\frac{J}{4}\sum_{I=1}^{N/3}[\sigma_{3,I}^{z}\sigma_{1,I+1}^{z}+D(\sigma_{3,I}^{x}\sigma_{1,I+1}^{y}-\sigma_{3,I}^{y}\sigma_{1,I+1}^{x})],
\end{equation}
where $\sigma_{j,I}^{\tau}$ represents the $\tau$ component of the classical Pauli matrix at site $j$ of the block labeled by $I$.

The two degenerate ground state of the corresponding $I$-th block are
 \begin{equation}
\begin{aligned}
|\varphi_{0}\rangle=\frac{1}{\sqrt{2q(1+q)}}[2D|100\rangle+i(1+q)|010\rangle-2D|001\rangle],\\
|\varphi_{0}^{'}\rangle=\frac{1}{\sqrt{2q(1+q)}}[2D|110\rangle+i(1+q)|101\rangle-2D|011\rangle],\\
\end{aligned}
\end{equation}
where $q=1/\sqrt{1+8D^{2}}$. The two lowest eigenstates can used to construct projection operator
 \begin{equation}
P=\prod_{i=1}^{N/3}P^{L}=\prod_{i=1}^{N/3}(|\Uparrow\rangle _{L}\langle \varphi|+|\Downarrow\rangle _{L}\langle \varphi|),
\end{equation}
where $\langle\Uparrow|, \langle\Downarrow|$ are renamed states of each block to
represent the effective site degrees of freedom. $|\varphi\rangle$ is the ground state $|\varphi\rangle_{0}$ or $|\varphi\rangle_{1}$. So using above method, we could get the effective Hamiltonian for this model.
\begin{equation}
H_{eff}=P^{\dagger}HP=H^{0}_{eff}+H^{1}_{eff}=P^{\dagger}H^{B}P+P^{\dagger}H^{BB}P\\
=\frac{J'}{4}[\sum_{i=1}^{N/3}\sigma_{i}^{z}\sigma_{i+1}^{z}+D'(\sigma_{i}^{x}\sigma_{i+1}^{y}-\sigma_{i}^{y}\sigma_{i+1}^{x})].
\end{equation}
where
\begin{equation}\label{eq11}
J'=J(\frac{1+q}{2q})^{2},~~~
D'=\frac{16D^{3}}{(1+q)^{2}}.
\end{equation}

\subsection{The critical behavior of quantum coherence}
\begin{figure}[!h]
\centering
\includegraphics[scale=0.5]{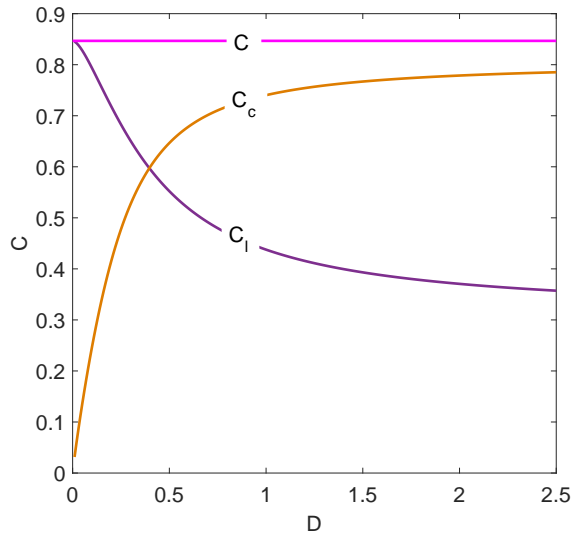}
\caption{Total coherence $C$, local coherence $C_{l}$, and collective coherence $C_{c}$ of Ising model with DM interaction as the function of $D$.}
\label{fig7}
\end{figure}
In Fig.~\ref{fig7}, we show the dependence of quantum coherence of Ising chain with DM interaction. The computational values of $C_{c}$, $C_{l}$, and $C$ also observe trade-off relation. These results prove the effectiveness that quantum coherence can be decomposed into various contributions.

\begin{figure}[!h]
\centering
\includegraphics[scale=0.5]{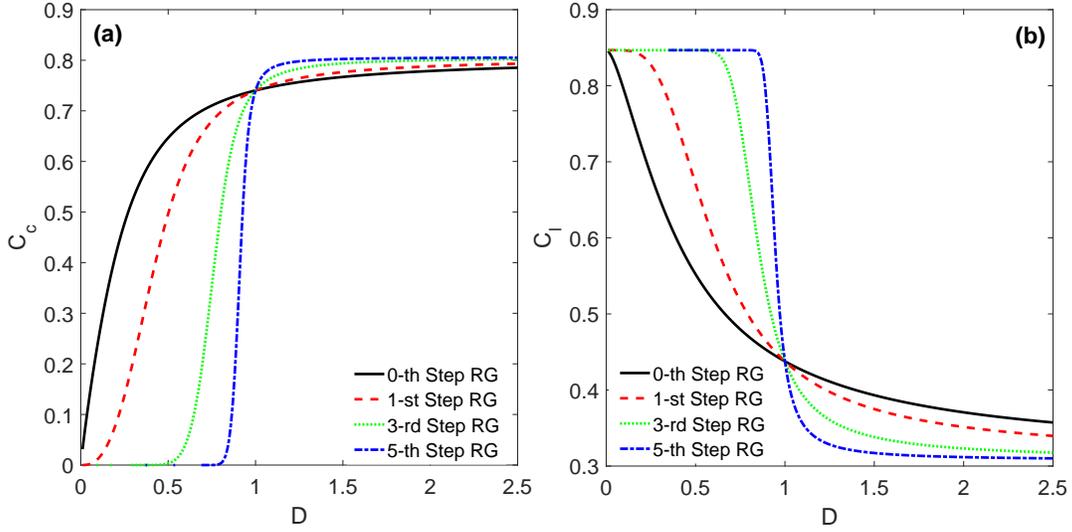}
\caption{Collective coherence $C_{c}$ and local coherence $C_{l}$ of Ising model with DM interaction as the function of $D$ at different QRG iteration steps.}
\label{fig8}
\end{figure}

In Fig.~\ref{fig8}, the results of collective coherence $C_{c}$ and local coherence $C_{l}$ after n-th QRG steps is depicted as a function of DM interaction. All curves cross at the critical point $D_{c}=1$. The collective coherence $C_{c}$ and local coherence $C_{l}$ develops its saturated values in both sides of the critical point. By increasing the size of the model, i.e. after QRG iterations, the collective coherence $C_{c}$ switches suddenly from  zero for $D<1$ to 0.806 for $D>1$, while local coherence $C_{l}$ changes suddenly from  0.846 for $D<1$ to 0.3105 for $D>1$.

\subsection{Non-analytic behavior and scaling behavior}

\begin{figure}[!h]
\centering
\includegraphics[scale=0.5]{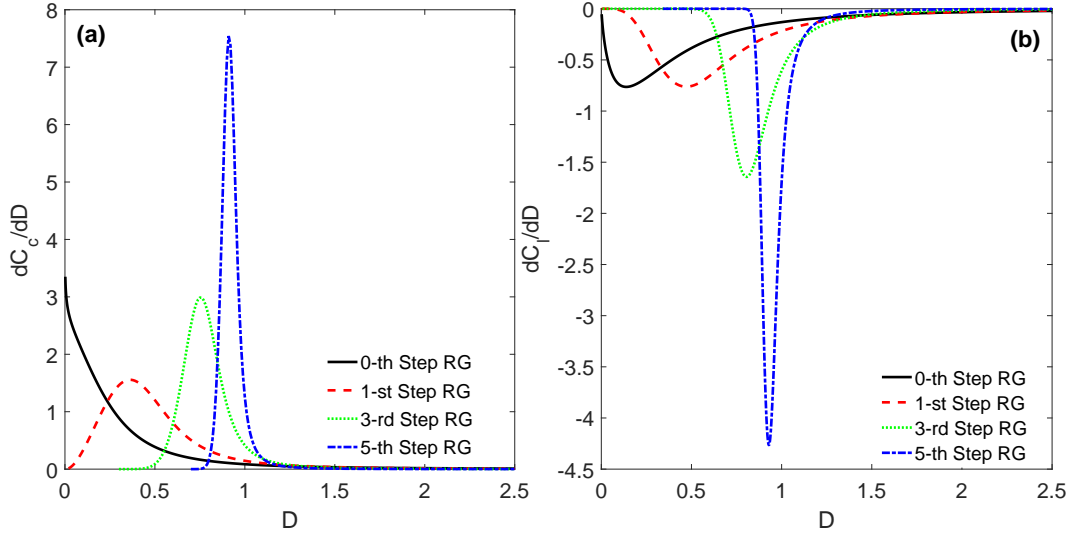}
\caption{The first derivative of collective coherence $C_{c}$ and local coherence $C_{l}$ of Ising model with DM interaction as the function of $D$ at different QRG iteration steps. }
\label{fig9}
\end{figure}

In order to get more details about the quantum critical behaviour of Ising chain with DM interaction, we investigate the variation of the first derivative of $dC_{c}/dD$ and $dC_{l}/dD$ as a function of DM interaction, as is shown in Fig.~\ref{fig9}. From the figure, we can find that there is a sharp peak for each QRG step with position $D$ that tends to $D_{c}=1$. The first derivative of the collective coherence $C_{c}$ and local coherence $C_{l}$ diverges near the critical point as the size of the model becomes larger. From the figure, it is noted that the singular property of the quantum coherence becomes more pronounced at the thermodynamic limit \cite{leila}.

\begin{figure}[!h]
\centering
\includegraphics[scale=0.5]{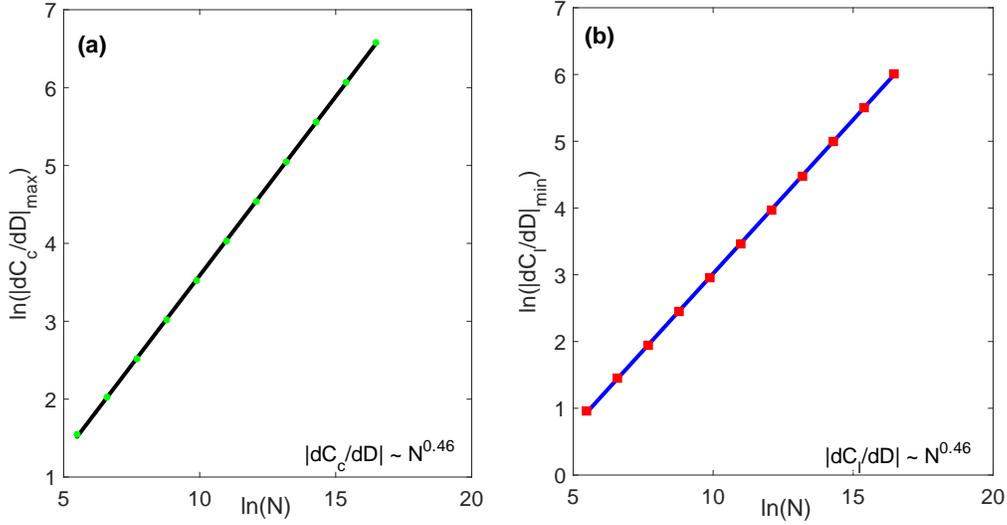}
\caption{The scaling behavior of $\ln |dC_{c}/dD|$ and $\ln |dC_{l}/dD|$ in terms of system size $\ln(N)$.}
\label{fig10}
\end{figure}

Accordingly, the positions of the maximum and the minimum of $dC/dD$ with the scale of the model increasing are given in Fig.~\ref{fig10}. It can be found that they both show a linear behavior in the system. The scaling law for this behavior is $|dC_{c}/dD|_{min}\sim N^{0.46}$ and $|dC_{l}/dD|_{max}\sim N^{0.46}$. These properties justify that $\theta$ is the reciprocal of the correlation length exponent $\nu$ near the critical point, i.e., $\theta=1/\nu$ \cite{mafw}.

\subsection{Monogamy of quantum coherence}
\begin{figure}[!h]
\centering
\includegraphics[scale=0.5]{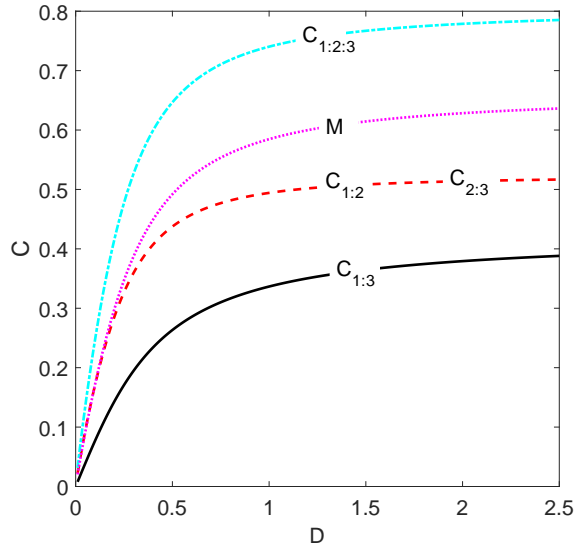}
\caption{The monogmay relation of Ising model with DM interaction as the function of $D$.}
\label{fig11}
\end{figure}
We have seen that quantum coherence can decompose into collective and local parts. Furthermore, it was recognized that a tripartite system $\rho_{123}$ also can decompose in a bipartite fashion \cite{radhakrishnan}
\begin{equation}
C_{123}<C_{1}+C_{2}+C_{3}+C_{2:3}+C_{1:23}.
\end{equation}

It is naturally arrive the concept of monogamy of quantum coherence. In a tripartite system, sharing quantum cherence
between several parties is restricted by the monogamy relation.
\begin{equation}
M=\sum_{n=2}^{N}C_{1:n}-C_{1:2,\ldots,N}.
\end{equation}
which is polygamous for $M>0$ as the dominant quantum coherence is distributed in a bipartite coherence. For $M\leq0$, it obey monogamous due to the multi-body coherence that is essential.

Now we discuss the behavior of monogamy coherence with the increase of DM interaction. In Fig.~\ref{fig11} we calculate Eq. (23) for the Ising model with DM interaction. The bipartite coherence of $\rho_{1:2}$, $\rho_{1:3}$, $\rho_{2:3}$, and tripartite coherence of $\rho_{1:2:3}$ are shown in the figure. We find that $M>0$ for all the DM interaction $D$, hence, the system is strictly polygamous in this case.

\section{\label{sec:level4}CONCLUSION}
In summary, we have use a coherence measure with metric nature to detect quantum phase transition by implementation QRG method.
This approach allows us to capture and explicitly derive the quantum critical point of block spin systems. We get the coherence behavior of a large scale systems by dealing with a small block which enable it possible to get accurate results. We find that the local coherence and collective coherence, decomposed from multipartite coherence, both can detect quantum critical point. In order to study the critical behavior of the Ising model, the evolution of both local coherence and collective coherence through the renormalization of the system were investigated. The first derivative of the two measures shows a diverging behavior as the scale of the system becomes large. Furthermore, the divergence of coherence are accompanied by some scaling behavior at the quantum critical point. As the coherence will decompose into different parts, we can therefore study the multipartite monogamy relation in the system, and, hence, giving another way of understanding the nature of many-body coherence.

\begin{acknowledgments}
This work was supported by the Natural Science Foundation of Jiangsu Province (Grant No. BK20171397)).
\end{acknowledgments}

\end{document}